\documentclass[mathleft
]{an}
\usepackage{graphicx}
\usepackage{times}
\begin{document}

\Pagespan{789}{}
\Yearpublication{2006}%
\Yearsubmission{2005}%
\Month{11}%
\Volume{999}%
\Issue{88}%

\title{Spectral components in black hole X-ray binaries}

\author{J. Malzac\inst{1,2}\fnmsep\thanks{Corresponding author:
  \email{julien.malzac@irap.omp.eu}\newline}
}
\titlerunning{Spectral components in BHBs}
\authorrunning{J. Malzac}
\institute{
Universit\'e de Toulouse, UPS-OMP, IRAP, Toulouse, France
\and  
CNRS, IRAP, 9 Av. colonel Roche, BP44346, F-31028 Toulouse cedex 4, France
}

\received{30 May 2005}
\accepted{11 Nov 2005}
\publonline{later}


\abstract{
 This paper summarises our current understanding of the spectral continuum components observed in black hole X-ray binaries. The consequences for theoretical models are discussed with an emphasis on the constraints set by observations on the nature of the X-ray corona in different spectral states.}

\maketitle

\section{Introduction}

Accreting black holes in X-ray binary systems (hereafter BHBs) produce radiation over the whole electromagnetic spectrum. From the radio to IR band, non-thermal emission is often detected and usually associated to synchrotron emission from very high energy particles accelerated in jets (see e.g. Chaty et al. 2003; Gandhi et al. 2011). The same particles can occasionally produce detectable emission in the GeV range (Fermi LAT Collaboration, 2009; Tavani et al. 2009; Malyshev et al. 2013). The thermal emission from the accretion disc can be studied in the optical to soft X-ray bands. In the hard X-ray domain (above a few keV) the emission is dominated by a non-thermal component  which nature and origin is strongly debated, and which is associated to the emission of a hot and tenuous  plasma located in the direct environment of the black hole. This plasma is generically called the "corona" drawing analogy from the solar corona, although the formation of the black hole corona, its power feeding and emission processes are most likely very different from that of the Sun.  

Most BHBs are transient X-ray sources detected during outbursts lasting from a few month to a few years. During outbursts their luminosity can increase by many orders of magnitude to reach values close to the Eddington limit ($L_{\rm E}\simeq 10^{39}$ erg/s for a 10 $M_{\odot}$ black hole) before going down, back to quiescence. These sources therefore constitute a unique laboratory to investigate how the physics of the accretion and ejection depends on the mass accretion rate onto the black hole. During these outbursts, not only the luminosity, but also the broad band spectral shape change drastically as a source evolves through a succession of X-ray spectral states, showing very different spectral (and timing) properties. Nevertheless those X-ray spectral states can all be described in terms of the same spectral components arising from the thermal accretion disc, X-ray corona and jets. The aim of this paper is to summarise the current understanding of these three components.

\section{Spectral states}

\begin{figure} 
\includegraphics[width=\linewidth]{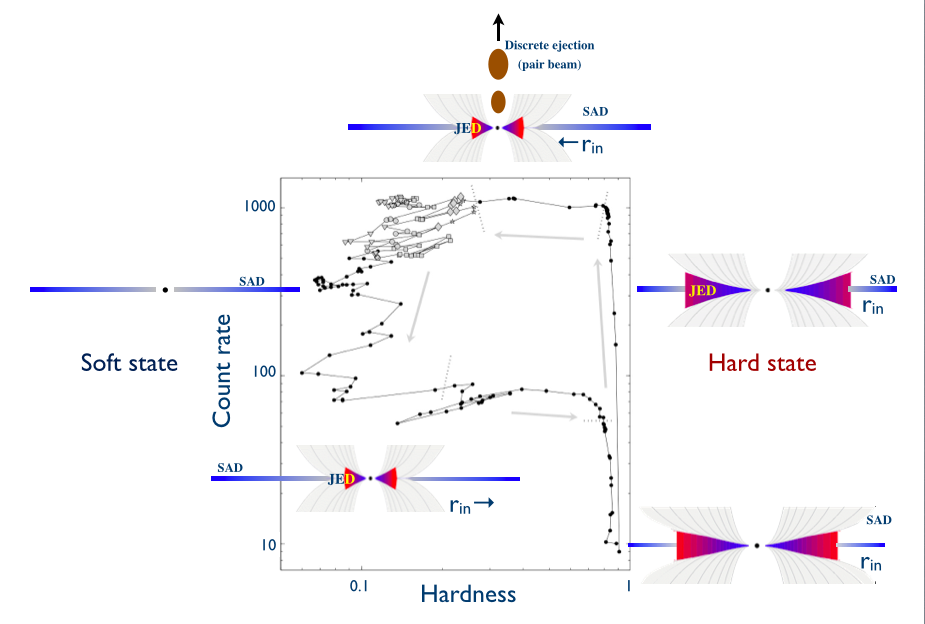} 
\caption{Typical track followed by the BHB GX339-4 in the HID diagram during an outburst.  The various sketches illustrate a possible scenario for the evolution of the geometry of the accretion flow in the context of a truncated disc model in which the central hot accretion flow takes the form of a jet emitting disc (Ferreira et al. 2006; Petrucci et al. 2010)} 
\label{fig:fig1} 
\end{figure}

\begin{figure*}
\center{\includegraphics[width=0.8\textwidth]{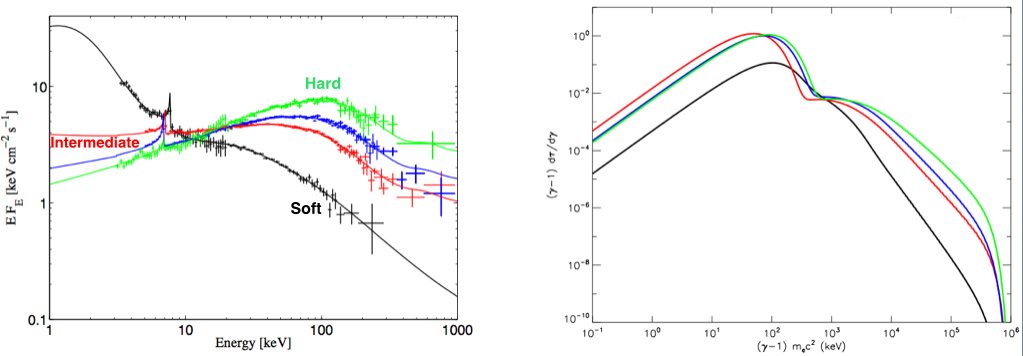}}
\caption{Left: Joint INTEGRAL/JEM-X, IBIS and SPI energy spectra of Cyg X-1 during four different spectral states fitted with the magnetised hybrid thermal-non-thermal Comptonisation model BELM. Right: Energy distribution of the Comptonising  electrons obtained in the best fit models of the left-hand side panel. These fits set an upper limit on the amplitude of the magnetic field in the X-ray corona at about $10^{5}$ G  in the harder states and $10^{7}$ G in softer states. From Del Santo al. 2013.}\label{fig:fig2}
\end{figure*}

The spectral evolution during outburst is usually studied using Hardness Intensity Diagrams (HID). One of such diagrams is shown in Fig.~\ref{fig:fig1}.
There are two main stable spectral states, namely the soft and the hard state, corresponding respectively to the left and right hand side vertical branches of the HID (see Fig.~\ref{fig:fig1}). The other spectral states are mostly short-lived intermediate states associated to transitions between the two main spectral states. Some X-ray spectra of Cyg~X-1 in its soft (in black), hard (green) and intermediate (blue and red) states are shown in Fig.~\ref{fig:fig2}.

The soft spectral state is observed at luminosity levels ranging approximately from $10^{-2}$  to a few  $0.1 L_{\rm E}$. In this state the high energy emission is dominated by the soft thermal multi-blackbody disc emission peaking around 1 keV, for this reason this state is also called 'thermal dominant' by some authors (Remillard \& McClintock 2006). The intense thermal radiation and hot disc temperature ($\sim$ 1 keV) is consistent with that of a standard  geometrically thin, optically thick accretion disc (Shakura \& Sunyaev 1973) extending down very close to the black hole. The coronal emission  is usually very weak forming a non-thermal power-law tail  above a few keV. The geometry of the corona is unconstrained in this state but  it is generally assumed to be constituted of small-scale magnetically active regions located above and below the accretion disc (e.g. Galeev, Rosner \& Vaiana 1979). In these regions the energetic electrons of the plasma up-scatter the soft X-ray photons coming from the disc into the hard X-ray domain.  
Due to the weakness of the non-thermal features, the soft state is perfect to test accretion disc models and measure the parameters of the inner accretion disc. In particular the detailed  model fitting of the thermal emission of the accretion disc indicate that the inner radius of the disc is a constant and is independent on the luminosity of the system (Gierli{\'n}ski \& Done 2004). This constant inner radius is believed to be located at the Innermost Stable Circular Orbit (ISCO) that is predicted by the theory of general relativity and below which the accreting material must fall very quickly across the event horizon. As the size of the ISCO is very sensitive to the spin of the black hole this offers an opportunity to constrain the spin of black holes in X-ray binaries. This may also be used to validate the spin measurements made using the relativistic  iron $K_{\alpha}$ line profile which remain the only method that can be used in AGN. This is a difficult task because measuring the disc inner radius accurately  requires very detailed disc emission model taking into account the general relativistic effects as well as non-blackbody effects through detailed disc atmosphere models (Davis et al. 2005, 2006). This method also requires the knowledge of the  distance and inclination of the system.  Nevertheless the most recent  BHBs spin estimates obtained from disc continuum  and  line fitting are converging (see Middleton 2015 for a recent review of these issues).  Although the soft state is associated to strong Fe absorption lines indicative of a disc wind (Ponti et al. 2014), there is no evidence so far of any relativistic jet component in soft state (see e.g. Russell et al. 2011).

The hard state is observed at all luminosities up to a few 0.1 $L_{E}$. In this state the emission from the accretion disc is much weaker and barely detected, the inferred temperature of the inner disc is also lower ($T_{\rm in}\sim$0.1 keV) than in soft state state (Cabanac et al. 2009; Dunn et al. 2011; Plant et al. 2015). The X-ray emission is dominated by a hard power-law with photon index $\Gamma$  in the range 1.5--2.1, and a high energy cut-off around 50-200 keV. This kind of spectra is very well represented by Thermal Comptonisation models (Sunyaev \& Titarchuk 1980) with Thomson depth 1-3 and electron temperatures in the range 20-200 keV. The amplitude of the reflection component due to illumination of the disc by the corona is usually small ($R< 0.3$). 

 The measurements of the hard state X-ray spectrum put interesting contraints on the geometry of the corona in the hard state.  For instance, an isotropic corona made of active regions located above the disc (as that envisioned for the soft state) would produce significantly stronger reflection features ($R\sim 1$ ) which may also be enhanced by light bending. In addition, most of the illuminating radiation would be  absorbed in the disc (typical disc X-ray albedo is $\sim 0.1$) and reemitted at lower energy in the form of nearly thermal radiation. This geometry would thus imply a strong thermal component with an observed flux that would be at least comparable to that of of the corona.  
 If the corona has a significant Thomson depth, as inferred by the observations ($\tau\ge 1$) and is radially extended above the disc, the reflection and reprocessing features might be smeared out by Compton scattering (Petrucci et al. 2001). However detailed Monte-Carlo simulations have shown that in this case  the strong reprocessed emission from the disc illuminating the corona would then cool down its electrons (via inverse Compton) to temperatures that are much smaller than observed (Haardt \& Maraschi 1993; Haardt et al. 1994; Stern et al. 1995, Malzac et al. 2001) and then much steeper X-ray spectra would be observed.  
 In fact, the high electron temperature and weakness of the reflection/and disc features suggest that the corona and the disc see each other with a small solid angle.

In this context, a geometry that is favoured is that of the truncated disc model, where the accretion disc does not extend down to the ISCO, but is truncated at some larger distance from the black hole. As the disc does not extend deeply into the gravitational potential of the black hole, its temperature is lower. As a consequence the blackbody emission is much weaker than in soft state. The corona is constituted of a hot geometrically thick accretion flow that fills the inner hole of the accretion disc (Esin, McClintock \& Narayan 1997; Poutanen, Krolik \& Ryde 1997). This hot flow Comptonises the soft photons from the disc and/or internally generated synchrotron photons to produce the hard X-ray continuum. The outer disc receives little illumination from the corona and produces only weak reflection and reprocessing features. This scenario explains qualitatively many of  the spectral and timing properties of BHBs in the hard state, such as the correlation between X-ray spectral slope and reflection amplitude  (Zdziarski, Lubi{\'n}ski \& Smith 1999) or QPO frequencies simply by assuming that the disc inner radius changes for instance with luminosity (see Done, Kubota, Gierli{\'n}ski 2007). The location of the disc truncation radius could be determined by a disc evaporation/condensation equilibrium (Meyer, Liu \& Meyer-Hofmeister 2000; Qiao \& Liu 2012). Fig.~\ref{fig:fig1} shows a sketch of the possible evolution of the geometry of the accretion flow during an outburst.

Observationally, however, due to the weak disc features the actual transition radius is very difficult to measure accurately. The most recent estimates suggest that in the bright hard state, the inner disc is actually at a few gravitational radii at most (Miller et al. 2015; Parker et al. 2015, Fabian et al. 2015). These results question whether the disc is actually truncated at all. But there is evidence that the disc recedes at low luminosity and the truncation radius could be located much farther out (see e.g. Plant et al. 2015).

 In the past decades there have been many theoretical studies on hot accretion flows which have been investigated both analytically and using numerical simulations (e.g. Igumenschev et al. 2003).  There are several possible hot accretion flow solutions involving either advection (ADAF, Ichimaru 1977; Narayan and Yi 1994; Abramowicz et al 1996), or outflows (ADIOS, Blandford \& Begelman 1999), or  convection (CDAF, Narayan et al. 2000) and/or jets (Petrucci et al. 2010) that could play the role of the hot inner corona.  
 However, it is worth noting that the only known hot flow solutions that can have a Thomson depth $\tau_T \ge 1$ as often observed in the bright hard state, requires strongly super-equipartition magnetic fields. Indeed, in these solutions the magnetic field pressure dominates and actually supports the hot flow. For a given vertical Thomson depth of the flow, the scale height can be higher than in standard flows reducing the flow density and therefore the plasma cooling rate, making hot solutions possible (Oda et al. 2010, 2012; Fragile \& Meier 2009). It is unclear however whether such strong magnetic fields can be generated and sustained in the hot flow without stopping the accretion by quenching the magneto-rotational instability that is believed to be at the origin of the viscosity. 
 
There are few alternatives to the truncated disc model. These include the outflowing corona model of Beloborodov (1999a). In this scenario the geometry is similar to that of the soft state (i.e. magnetically active coronal regions atop a thin disc extending down to the ISCO). The problem of the strong reprocessing features expected for this geometry is removed by assuming that the corona is moving away from the disc with a mildly relativistic velocity. Due to strong Doppler beaming effects the coronal emission is then strongly anisotropic and directed preferentially away from the disc. This model reproduces well the observed hard state spectra (Malzac, Beloborodov \& Poutanen 2001). The  correlation between X-ray spectral slope and reflection amplitude can be produced by varying the speed of the outflowing corona. The difficulty with this model is that the physics of this outflowing corona is not very well understood. An electron-positron pair plasma would be easily blown out at such speed by the radiation pressure from the disc (Beloborodov 1999b) but in the case of a normal ion electron plasma the acceleration mechanism is not specified. Moreover, even if the reprocessing features are weak due to the radiation anisotropy, the model requires most of the power to be dissipated in the corona and the intrinsic dissipation in the disc must be negligible. In other word the disc is essentially passive and most of the accretion power has to be transported from the disc to the corona by some unspecified mechanism.  Another  alternative would be for the hard X-ray continuum to originate from a jet rather than the accretion flow (see below). 

\section{Hard state jets}

\begin{figure}
   \includegraphics[width=\linewidth]{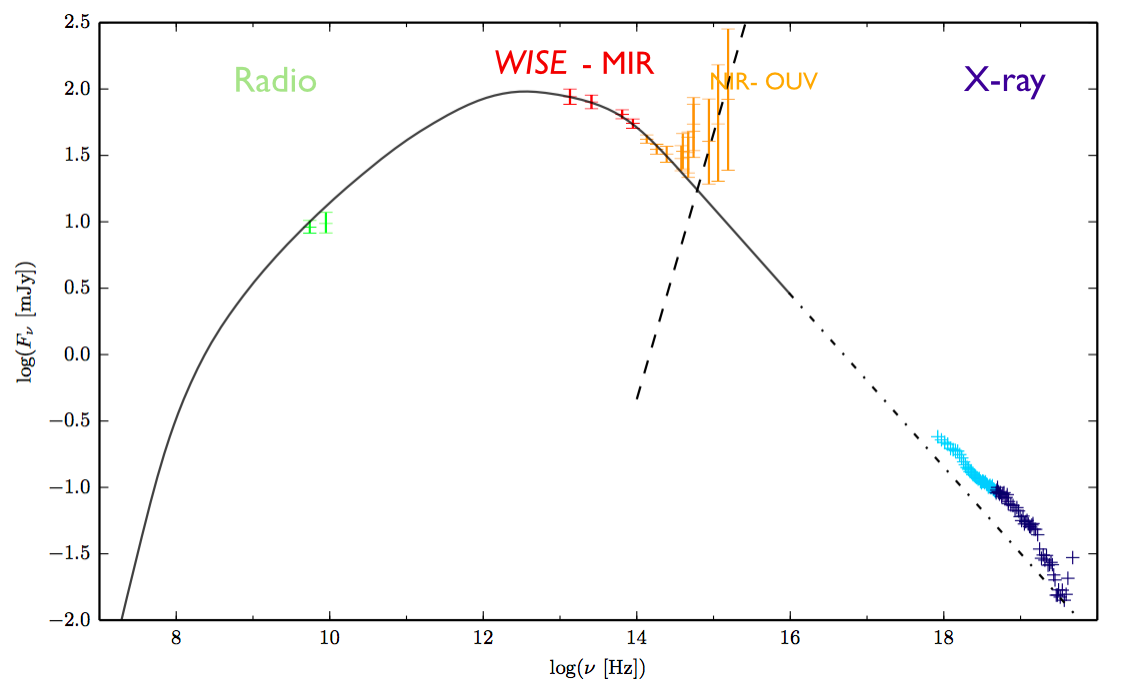}
 \caption{The SED of GX339-4 measured by Gandhi et al. (2011) compared to a synthetic jet SED  obtained from the internal shock model assuming that the jet Lorentz factor fluctuations have exactly the same Power Density Spectrum as the X-ray flux (from Drappeau et al. 2015)}
 \label{fig:fig3}
 \end{figure}

The hard state is associated with the presence of compact radio jets which are inferred from a strong radio emission. This radio flux can be several orders of magnitude higher than the upper limits obtained in soft state (e.g. Corbel et al. 2000; Corbel \& Fender 2002). In some cases, the jet structure is resolved (Stirling et al. 2001). The jet emission is characterised by a flat to nearly inverted radio spectrum interpreted as partially self-absorbed synchrotron radiation from  relativistic electrons in the jet. This flat emission spectrum  seems to extend a least to the far infrared where there is evidence for a spectral break where the slope the spectrum steepens. This break is believed to be due to the transition from partially self-absorbed to optically thin synchrotron (e.g. Corbel \& Fender 2002; Gandhi et al. 2011, see Fig.~\ref{fig:fig3}). This kind of Spectral Energy Distributions (SEDs)  are usually interpreted in terms of the conical jet emission model of Blandford \& K\"onigl (1973) or some elaboration upon it. Recent developments suggest that the acceleration of particles within the compact jet might be triggered by internal shocks driven by fluctuations of the ejection velocity and subsequent collisions at a range of distances from the jet (Malzac 2013, 2014). Interestingly these fluctuations of the jet velocity appear to be related to the variability of the accretion flow as inferred from the X-ray light curve (Drappeau et al. 2015). If confirmed this would offer a new way to probe the dynamics of the connection between accretion and ejection processes.

Another interesting point is that the slope of the optically thin synchrotron emission at frequencies above the IR break is both measured and theoretically expected to be hard: photon index $\Gamma \sim 1.6$  i.e. comparable to that observed in hard X-rays. Moreover in several sources, including GX 339-4, the extrapolation of the optically thin  IR/optical component into the X-ray range falls very close to the actual X-ray flux (see e.g. Fig.~\ref{fig:fig3}). This "coincidence" and also the fact that the hard state X-ray flux is tightly correlated to the radio flux (e.g. Corbel et al. 2013) led to the suggestion that the jet synchrotron emission could dominate the hard state X-ray spectral flux (Markoff et al. 2001).    
However this was later shown to be quite unlikely. Jets are know to be very radiatively inefficient, and the jet X-ray luminosity is expected to represent less than a few percent of the total kinetic power of the jet. Thus a large luminosity jump would be expected during state transition from radiatively inefficient jet to efficient disc dominated flow, but is not seen (Maccarone 2005). Also, despite having X-ray spectral and variability features very similar to BHBs, accreting neutron stars are clearly not jet dominated. At a given X-ray luminosity they have a much fainter radio jet. In the neutron star 4U0614+91, in which the IR break is clearly detected, the extrapolation of the IR synchrotron spectrum into the X-rays falls several orders of magnitude below the observed X-ray flux (Migliari et al. 2010). 

Other X-ray jet models assume that the jet constitutes the Comptonising cloud (e.g. Kylafis et al. 2008). In fact, the huge mass  ejection rate implied by the measured large Thomson depths would require a huge jet kinetic power, which again would predict a jump in X-ray luminosity at state transition. In Cyg~X-1 the required jet power would be much larger  than current observational estimates based on the measurements of $H_{\alpha}$ and [OIII]  measurements of the optical nebula surrounding the X-ray source (Malzac, Belmont \& Fabian 2009).
Note that although it is unlikely that the jet dominates the X-ray emission in hard state, a significant contribution to the X-ray flux is not ruled out. Moreover, the corona appears to be intrinsically related to the jet and could constitute the jet launching region. 

\section{Non-thermal particles} 

In the soft state there are indications (at least in the prototypical source Cyg X-1) that the  hard X-ray emission extends as a power law at least up to a few MeV (McConnell et al. 2002) . The absence of cut-off below 1 MeV indicates that the Comptonising electron distribution of the soft state corona cannot be purely thermal. Producing such gamma-ray emission through inverse Compton requires a power-law like distribution of electrons extending at least up to energies of order of 10-100 times the electron rest mass energy.  In the hard state, an excess with respect to a pure Comptonisation model is detected in all bright sources (e.g. Joinet et al. 2007; Droulans et al. 2010; Jourdain et al. 2012a)  which is also interpreted as the signature of a population of non-thermal electrons in the corona. 

These findings triggered the development of hybrid thermal/non-thermal Comptonisation models. In these models the Comptonising electrons have  a similar energy distribution in all spectral state i.e. a Maxwellian with the addition of a high energy power-law tail (Poutanen \& Coppi 1998; Coppi 1999).
These models  have been extremely successful at fitting the broad band high energy spectra BHBs. Fig.~\ref{fig:fig2} shows examples of INTEGRAL spectra fit with an hybrid model. 
The transition from mostly thermal (in hard state)  to mostly non-thermal  (in soft state) emitting electrons is understood in terms of the radiation cooling. As a source evolves towards the soft state the corona intercepts a much larger flux of soft photons from the disc. In this more intense soft radiation field the Compton emission of the hot electrons of the plasma is stronger. They radiate their energy faster. This makes the electron temperature significantly lower in the softer states.  As a consequence the Compton emissivity of the thermal particles is strongly reduced in the soft state and the emission becomes dominated by the higher energy non-thermal particles. In addition the Thomson depth of thermal electrons is found to be smaller in soft state, possibly because most of the material in the corona has condensed into the disc or is ejected during state transition. This further decrease the luminosity of the Maxwellian component of the plasma, making it barely detectable in soft state. Attempts to model the spectral evolution during spectral state transitions have confirmed that the huge change in the flux of soft cooling photon from the disc illuminating the corona drives the spectral changes of the corona (Del Santo et al. 2008).

The most recent version of the model also includes the radiative effects of magnetic field on the lepton energy distribution (Belmont, Malzac \& Marcowith 2008; Vurm \& Poutanen 2009).  Internally generated synchrotron photons (typically in the optical/UV range) constitute a source of seed  photons for the Comptonisation process. In addition, the process of Synchrotron self-absorption  provides an efficient coupling between leptons which can quickly exchange energy by rapid emission and absorption of synchrotron photons leading to very fast thermalisation of the lepton distribution on time-scales comparable to the light crossing time. In fact, it is not necessary to assume the presence of thermal Comptonising electrons in the first place. The heating mechanism could be purely non-thermal e.g. accelerating all electrons into a power-law energy distribution. The thermalising effects of the so-called synchrotron boiler (in addition to Coulomb collisions) naturally leads to an hybrid thermal/non-thermal particle energy distribution (Malzac \& Belmont 2009; Poutanen \& Vurm 2009). The detailed comparison of this model with observations has brought interesting constraints on the magnetic field. In the hard state in particular, the conclusion was that either the magnetic field is strongly sub-equipartition (which would be in contradiction with theoretical models involving a magnetically dominated accretion flow, or accretion disc corona atop the disc), or,  the MeV excess is produced in a region that is spatially distinct  from that producing the bulk of the hard X-ray radiation (Del Santo et al. 2013).

In the latter case,  the accretion flow could for instance, be constituted of a truncated accretion disc surrounding a central magnetically dominated hot accretion flow responsible for the thermal Comptonisation component, while active coronal regions above and below the outer disc may produce the mostly non-thermal Compton emission (i.e. the hard state MeV excess, Malzac 2012). In the soft state the disc inner radius moves inwards until it reaches the ISCO and the emission becomes gradually dominated by the non-thermal corona. Another interesting possibility is suggested by INTEGRAL  measurements showing  that the hard state MeV tail of Cyg~X-1 is strongly polarised (at a level of 70 $\pm$30 percent), while the thermal Comptonisation emission at lower energy is not (Laurent et al. 2011; Jourdain et al. 2012b). The only plausible mechanisms for the very high levels of polarisation seems to be synchrotron emission in a highly coherent magnetic field. A natural explanation would be that the MeV excess is in fact a contribution from the jet (but see Romero, Vyero \& Chaty 2014). Conventional jet models can produce such MeV component but require some fine tuning and quite extreme acceleration parameters (see Zdziarski et al. 2014)

\section{Spectral evolution in the Hard state}

All the above considerations about the spectral states  are limited to the high luminosity levels (typically $>0.01 L_{E}$) for which a detailed spectral analysis is possible.  At lower luminosity, the variations of spectrum can be detected by measuring the hardness ratios, or the photon index of the X-ray spectrum.  In the hard state the photon index depends on luminosity. Below a percent of $L_{\rm E}$ the spectrum is harder when brighter, while above this critical luminosity the trend turns to the opposite and the spectrum is softer when brighter (Wu \& Gu 2008). This can be understood qualitatively in terms of the truncated disc model, if one assumes that the inner disc radius decreases monotonically with luminosity. Despite the changes in inner disc radius, the dimension of the coronal region producing most of the X-ray emission is limited to the direct environment of the black hole where most of the gravitational energy of the accreting gas can be released. At low luminosities, the accretion disc is very far from the black hole and the X-ray corona does not receive much soft seed photons from the disc. The seed photons for Comptonisation are then provided mostly by synchrotron emission of the coronal electrons. Applying the standard scalings of the hot accretion flow  density and plasma magnetic field and luminosity with mass accretion rate, one can show that  the slope of the accretion flow should harden when the luminosity increases (see e.g. Nied{\'z}wiecki et al. 2014, 2015). It is possible that around  the critical luminosity of $10^{-2}  L_{E}$, the inner disc approaches the coronal X-ray emitting region sufficiently to provide a significant amount of soft photons to the Comptonisation process. Then, as the inner disc radius approaches the ISCO, the flux of soft disc photons illuminating the corona increases. This increases the cooling rate of the Comptonising electrons and the temperature of the plasma gradually decreases, leading to a softening of the spectrum (Sobolewska et al. 2011).  Alternatively the softening at high luminosity could be due to the formation of an inner disc ring illuminating the corona (Qiao \& Liu 2013).
An interesting question is to understand what happens at the lowest luminosities. It turns out that for luminosities below $10^{-4} L_{E}$ the spectral slope saturates around $\Gamma\simeq 2.1$ and becomes independent of luminosity (Plotkin et al. 2013). This suggests that we observe the emission from non-thermal particles (i.e. power-law) rather than thermal Comptonisation. Unlike that of the jet, the radiative efficiency of hot accretion flows decreases toward lower mass accretion rates/luminosities. Below some luminosity the jet synchrotron emission may dominate the X-ray emission (Yuan \& Cui 2005). Recent NuSTAR results indicate this could be the case in V404 Cyg  (Rana et al. 2015). Note that if this is correct, the transition from accretion flow to jet dominated source is expected to produce a change in the slope of the X-ray vs radio correlation around X-ray luminosities  $\sim 10^{-4} L_{E}$, which is not observed.  Another possibility could be that, because the mass accretion rate vanishes, the Thomson depth of the Maxwellian component of the plasma becomes very small. As a consequence, the Compton emissivity of thermal particles in the corona decreases. The emission becomes dominated by the population of non-thermal particles in the same way as during  hard to soft spectral state transition.  

\section{Conclusion}

Although the truncated disc  scenario  explains qualitatively many of the observed properties of black hole binaries, in the  best documented bright hard state sources none of the 'usual' accretion flow models really fit the data. In fact the nature of the corona in all spectral states is still uncertain. Spectral fits with magnetised Comptonisation models suggest that the structure of the corona is more complex (multi-zone) than previously envisioned, or that the jet contributes to the observed MeV emission in the hard state.
 Jets could contribute to the X-ray emission but are unlikely to represent a dominant Component in the hard state.  Nevertheless the properties of the jet appears to be strongly related to that of the inner accretion flow. In particular recent modelling of the jet emission suggest that the radio-IR jet SED might be related to the X-ray timing properties.  The usual spectral state classification regards the behaviour of the sources at relatively high luminosity  for which the photon statistic is good enough to perform detailed spectral analyses.  The observed variation of the X-ray hardness indicate that there are additional radiative transitions occurring at lower luminosities. 



\acknowledgements
The author acknowledges funding from the french Research National Agency: CHAOS project ANR-12-BS05- 0009 (http://www.chaos-project.fr). 


\end{document}